\documentclass[prx,twocolumn, amsmath,amssymb,
reprint,%
author-year,%
author-numerical,%
dvipdfmx
]{revtex4}

\usepackage{graphicx}
\usepackage{bm}
\usepackage{color}

\begin{document}


\title{Metastability relationship between two- and three-dimensional crystal structures: \\
A case study of the Cu-based compounds}
\author{Shota Ono}
\email{shota\_o@gifu-u.ac.jp}
\affiliation{Department of Electrical, Electronic and Computer Engineering, Gifu University, Gifu 501-1193, Japan}

\begin{abstract}
Some of the three-dimensional (3D) crystal structures are constructed by stacking two-dimensional (2D) layers. It remains unclear whether this geometric concept is related to the stability of ordered compounds and whether this can be used to computational materials design. Here, using first principles calculations, we investigate the dynamical stability of copper-based compounds Cu$X$ (a metallic element $X$) in the B$_h$ and L1$_1$ structures constructed from the buckled honeycomb (BHC) structure and in the B2 and L1$_0$ structures constructed from the buckled square (BSQ) structure. We demonstrate that (i) if Cu$X$ in the BHC structure is dynamically stable, those in the B$_h$ and L1$_1$ structures are also stable. Although the interrelationship of the metastability between the BSQ and the 3D structures (B2 and L1$_0$) is not clear, we find that (ii) if Cu$X$ in the B2 (L1$_0$) structure is dynamically stable, that in the L1$_0$ (B2) is unstable, analogous to the metastability relationship between the bcc and the fcc structures in elemental metals. The total energy curves for Cu$X$ along the tetragonal and trigonal paths are also investigated. 
\end{abstract}

\maketitle

\section{Introduction}
\begin{figure}[bb]
\center
\includegraphics[scale=0.27]{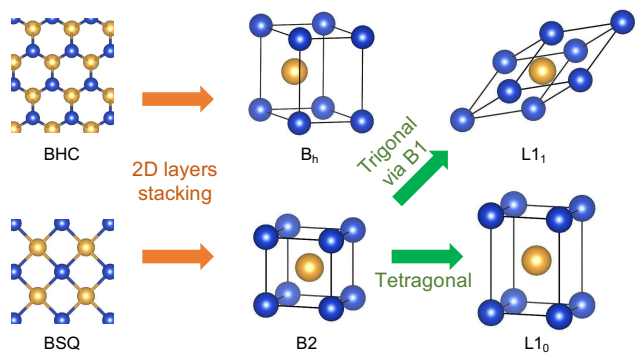}
\caption{Illustrations of the 2D (BHC and BSQ) and 3D structures (B$_h$, L1$_1$, B2, and L1$_0$) and its interrelationships: Stacking the 2D structures vertically produces the 3D structures and the tetragonal and trigonal distortions of the B2 structure yield the L1$_0$ and L1$_1$ structures, respectively.  } \label{fig_crys} 
\end{figure}

Since the synthesis of many atomically thin materials, the two-dimensional (2D) structure has been regarded as one of the metastable structures in materials science, as a result of which the database including several 2D structures can now be available \cite{ashton,choudhary,haastrup,feng}. Many 2D materials can be exfoliated from their three-dimensional (3D) counterparts, as the relation between the graphene and the graphite, in turn, implying that the 2D layers can be building blocks for constructing the 3D crystal structures. Recently, 2D CuAu has been synthesized experimentally \cite{zagler}, where it consists of the hexagonal Cu and Au monolayers, forming the buckled honeycomb (BHC) structure. By considering the 2D CuAu as a building block for the 3D structures, one can construct the B$_h$ and L1$_1$ structures with the ABAB and ABC stacking methods, respectively (see Fig~\ref{fig_crys}). However, the synthesis of CuAu in these structures have not yet been reported. 

The stability of the 2D metals has recently been studied in detail \cite{akturk,kochat,nevalaita,hwang,wang2020,ren}. By focusing on the 2D elemental metals, the author has demonstrated that the concept above (i.e., the 2D structure as a building block for the 3D structures) can hold by using first principles calculations: If the planar hexagonal (HX) structure is dynamically stable, then the BHC, the fcc, and/or the hcp structures are also stable \cite{ono2020}. Po in the square lattice structure is dynamically stable \cite{ono2020_Po}, as a counterpart of Po in the simple cubic structure. More recently, the author has also demonstrated that if a compound in the B$_h$ structure has been synthesized experimentally, that in the BHC structure is dynamically stable \cite{ono2021}. These studies motivate us to study the metastability relationship between the 2D and 3D compounds in detail. 

The crystal structures of Cu-based compounds have been studied for many years since the discovery of CuAu compounds in the L1$_0$ structure. For the binary compounds of Cu$X$, where $X$ is an element in the periodic table, several phases have been synthesized experimentally: CuBe \cite{CuBe}, CuPd \cite{CuPd}, CuSc \cite{CuSc}, CuY \cite{CuY}, CuZn \cite{CuZn}, and CuZr \cite{CuZr} in the B2 (CsCl-type) structure, CuCl, CuBr, and CuI \cite{copper_halides} in the B3 (zincblend-type) structure, CuAu \cite{CuAu} in the L1$_0$ structure, and CuPt \cite{CuPt} in the L1$_1$ structure. These structures are interrelated with each other by the following deformations: The tetragonal Bain distortion elongating the $c$ axis transforms the B2 into the L1$_0$ structure (see Fig~\ref{fig_crys}); the trigonal distortion elongating the (111) axis transforms the B2 structure into, via the B1 (NaCl-type) structure, the L1$_1$ structure (see Fig~\ref{fig_crys}); and by shortening the interatomic distance along the (111) axis the B1 is transformed into the B3 structure. In this way, the existence of Cu$X$ in the B$_h$ structure has not yet been explored.

In this paper, we investigate the interrelationship of the dynamical stability between the 2D and the 3D structures by focusing on Cu$X$ with a metallic element $X$. By calculating the phonon density-of-states (DOS) from first principles, we demonstrate that if Cu$X$ in the BHC structure is dynamically stable, Cu$X$ in the B$_h$ and L1$_1$ structures are also stable. This indicates that the BHC structure serves as a building block for the B$_h$ and L1$_1$ structures, while in some compounds the dynamical stability of the B$_h$ and L1$_1$ structures relies on the coupling between the BHC structures along the $c$ axis. 

\begin{table*}
\begin{center}
\caption{The interrelationship of the dynamical stability between different structures. The ``stable'' structure has no imaginary frequencies within the phonon Brillouin zone. }
{
\begin{tabular}{lcll} \hline\hline
Property 1 & $\rightarrow$ & Property 2 & Example \\
\hline
fcc and hcp stable & $\rightarrow$ & bcc unstable & group 3, 4, 10, and 11 \cite{grimvall} \\  
bcc stable & $\rightarrow$ & fcc and hcp unstable & group 5 and 6 \cite{grimvall} \\  
HX stable & $\rightarrow$ & BHC stable & group 2, 11, and 12 \cite{ono2020} \\  
HX stable & $\rightarrow$ & fcc and hcp stable & group 2, 11, and 12 \cite{ono2020} \\  
BSQ stable & $\rightarrow$ & fcc and hcp stable & group 3, 7, 8, 9, and 10 and Al \cite{ono2020} \\  
B$_h$ synthesized & $\rightarrow$ & BHC stable & IrLi, LiPd, LiPt, and LiRh \cite{ono2021} \\  
BHC stable & $\rightarrow$ & B$_h$ and L1$_1$ stable & Cu$X$ with $X=$ Li, W, Co, Ni, Cu, Ag, Au, Zn, and Al (this work) \\  
B2 stable & $\rightarrow$ & L1$_0$ unstable & Cu$X$ with $X=$ K, Rb, group 2 and 3, and Pd (this work) \\  
L1$_0$ stable & $\rightarrow$ & B2 unstable & Cu$X$ with $X=$ Li, group 5 and 6, Mn, Tc, Fe, Ru, group 9, Ni, Pt, Cu, Au, \\
& & & Zn, and Ga (this work) \\  
\hline\hline
\end{tabular}
}
\label{table1}
\end{center}
\end{table*}

We also study the dynamical stability of Cu$X$ in the the buckled square (BSQ), B2, and L1$_0$ structures, where the latter two structures can be constructed by stacking the BSQ structures (see Fig~\ref{fig_crys}). However, we find it difficult to discuss the metastability relationship between the 2D and 3D tetragonal structures because most of Cu$X$ in the BSQ structure are unstable. Analogous to the instability of the bcc (fcc) elemental metals in the fcc (bcc) structure \cite{grimvall}, we can find that if Cu$X$ in the B2 (L1$_0$) structure is dynamically stable, that in the L1$_0$ (B2) structure is unstable. This is visualized by means of the tetragonal Bain path calculations. 

The metastability relationships between different crystal structures are summarized in Table \ref{table1}: The first two relationships between the fcc, hcp and bcc structures can be applicable to the stability of elemental metals \cite{grimvall}; the third is the relationship between the 2D elemental metals; the fourth and fifth are the relationships between the 2D and 3D elemental metals; and the others are the relationships between the 2D and 3D compounds. It has been shown that the $(m,n)$ Lennard-Jones crystals also satisfy the first and the fifth relationships when $m+n\ge 18$, i.e., the short-ranged potentials \cite{ono2021_LJ}.  

Among binary metallic phases, L1$_1$ structure is a quite rare structure, as has been pointed out in Ref.~\cite{nelson}. B$_h$ structure is also a rare structure, as observed in AlSn solid solutions \cite{kane_AlSn}. The present search based on the geometric concept predicts that many Cu$X$ in the B$_h$ and L1$_1$ structures are dynamically stable, which supplements the formation energy analysis in the search of the rare structures \cite{nelson}. 

The geometry-based approaches have been proposed in computational materials science. Hart et al. have generated superstructures as a derivative of a parent crystal structure \cite{hart2008}. In contrast, Kolli et al. have mapped complex crystal structures onto several parent structures \cite{kolli2020}. They have pointed out that 15 of the top 20 parent structures can be generated by vertically stacking simple 2D structures including the close-packed, honeycomb, kagome, square, and more complex structures. As an extension of the present investigation, these will be candidate structures for creating the metastability database of the ordered compounds $XY$. 


\begin{figure*}[tt]
\center
\includegraphics[scale=0.45]{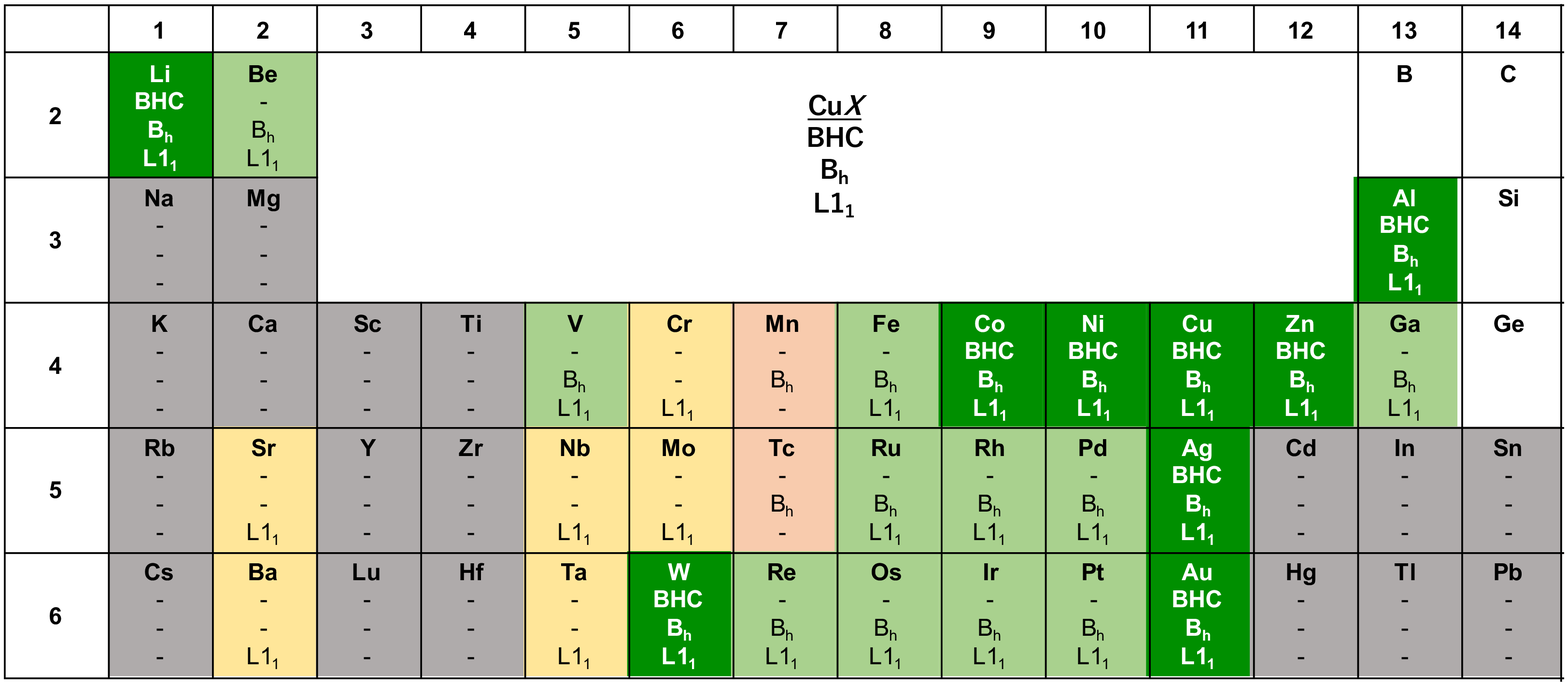}
\caption{Periodic table for the Cu$X$ compounds in the BHC, B$_h$, and L1$_1$ structures. The dynamically stable structure $j$s are indicated. Cu$X$ with $X=$ Ca, Sr, Ba, Sc, Y, and Lu has the B1 structure rather than the L1$_1$ structure as a dynamically stable or unstable structure (see the SM \cite{SM}). } \label{fig1} 
\end{figure*}

\begin{figure*}[tt]
\center
\includegraphics[scale=0.45]{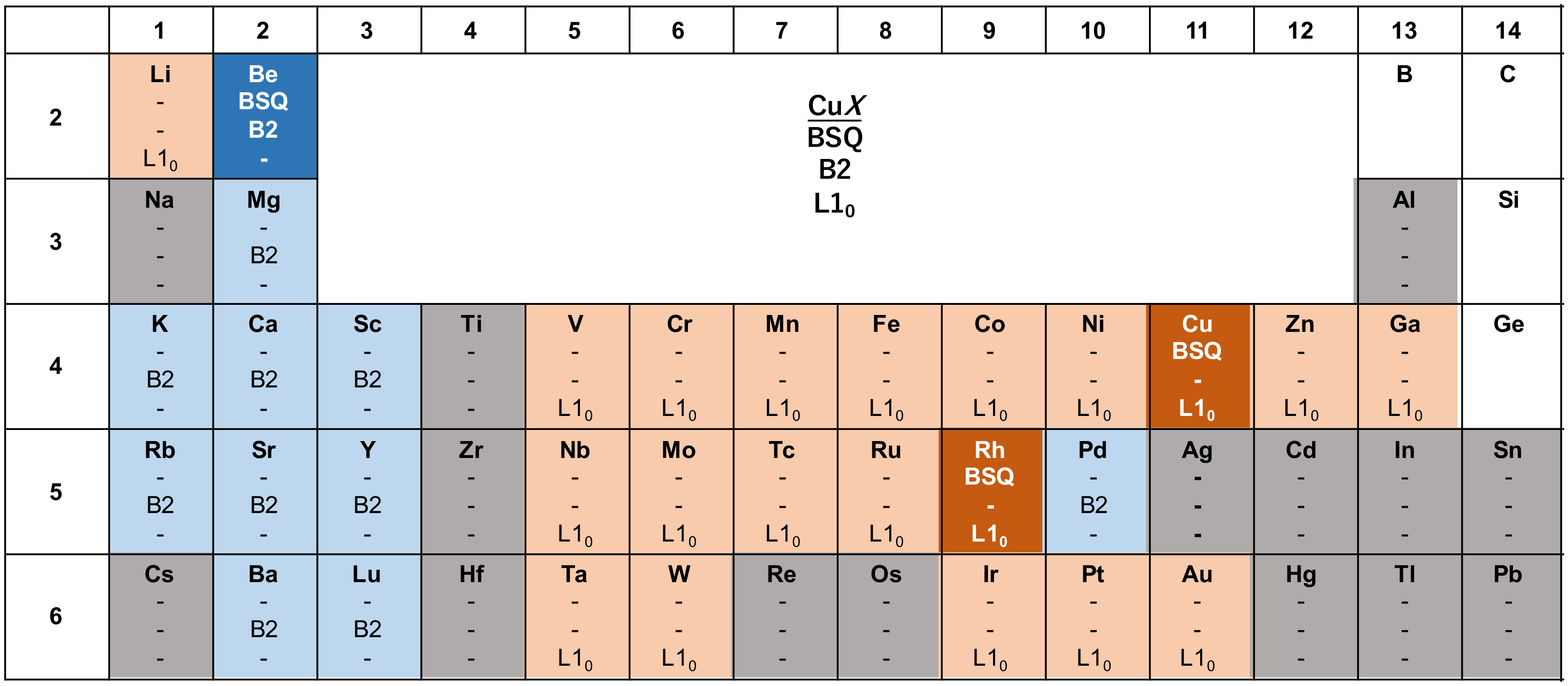}
\caption{Same as in Fig.~\ref{fig1} but for the Cu$X$ compounds in the BSQ, B2, and L1$_0$ structures. } \label{fig2} 
\end{figure*}

\section{Results and discussion}
\subsection{BHC, B$_h$, and L1$_1$}
\label{sec:BHC_Bh_L1_1}
The resultant dynamical stability properties are summarized in the periodic table of Fig.~\ref{fig1}. The phonon DOS of Cu$X$ in the BHC, B$_h$, and L1$_1$ structures are provided in the Supplemental Materials (SM) \cite{SM}. For the case of $X={\rm Cu}$, the B$_h$ and L1$_1$ structures correspond to the hcp and fcc structures, respectively. The main finding is that if the BHC structure is dynamically stable, then the B$_h$ and L1$_1$ structures are also stable, as in the cases of $X=$Li, W, Co, Ni, Cu, Ag, Au, Zn, and Al (dark green). On the other hand, if the B$_h$ and L1$_1$ structures are unstable, the BHC structure is also unstable (gray). These indicate that the BHC structure serves as a building block for the B$_h$ as well as L1$_1$ structures. For some compounds (light green, light red, and yellow), only the B$_h$ and/or L1$_1$ structures are dynamically stable. This implies that the layer stacking vertically suppresses the instability against the out-of-plane vibrations in thin films, while the prediction of such a critical thickness is beyond the scope of the present work. 

\begin{figure*}[tt]
\center
\includegraphics[scale=0.45]{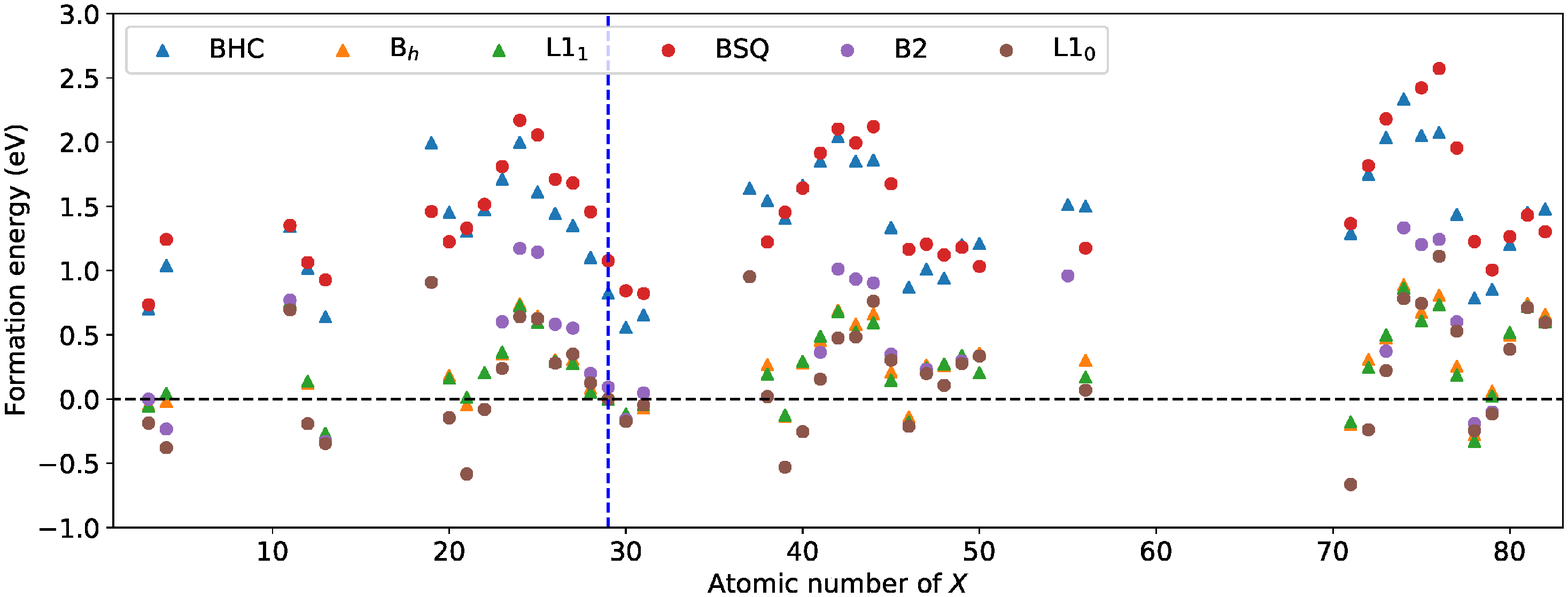}
\caption{The formation energy (per a unit cell including two atoms) of Cu$X$ in the BHC, B$_h$, L1$_1$, BSQ, B2, and L1$_0$. The vertical dashed line (blue) indicates the case of $X=$ Cu. } \label{fig_Eform} 
\end{figure*}

The formation energies $E_{j}({\rm Cu}X)$ defined as Eq.~(\ref{eq:Eform}) for the structure $j=$ BHC, B$_h$, and L1$_1$ structures (triangles) are plotted in Fig.~\ref{fig_Eform}. The Cu$X$ in the BHC structure have positive $E_j$. For the cases of $X=$ Li, Be, Sc, Y, Lu, Pd, Pt, Zn, Al, and Ga, the B$_h$ and/or L1$_1$ structures have negative $E_j$. As discussed in Refs.~\cite{wang,ono2021}, a negative value of $E_{j}$ is neither a sufficient nor necessary condition for the dynamical stability of the ordered compounds. In fact, although CuY and CuLu in the B$_h$ and L1$_1$ structures have negative $E_{j}$, CuY and CuLu in the B$_h$ and L1$_1$ structures are unstable (Fig.~\ref{fig1}). In addition, irrespective to the positive $E_{j}$, 9 of the 46 Cu$X$ in the BHC structure are dynamically stable. 



\subsection{BSQ, B2, and L1$_0$}
\label{sec:BSQ_B2_L1_0}
We next study the dynamical stability of Cu$X$ in the BSQ, B2, and L1$_0$ structures. The phonon DOS of these structures are also provided in the SM \cite{SM}. As shown in Fig.~\ref{fig2}, only CuBe, CuRh, and CuCu in the BSQ structure are dynamically stable, while most of Cu$X$ in the B2 or L1$_0$ structures are dynamically stable. Therefore, it seems to be difficult to find a useful relationship of the dynamical stability between the BSQ and the 3D structures. Interestingly, one can find that if Cu$X$ in the B2 structure is dynamically stable, that in the L1$_0$ structure is unstable, and vice versa: if Cu$X$ in the L1$_0$ structure is dynamically stable, that in the B2 structure is unstable. This is analogous to the metastability relationship between the bcc and the fcc structures in the elemental metals listed in Table \ref{table1}. This will be rationalized by the tetragonal Bain path calculations below. 

As shown in Fig.~\ref{fig_Eform} (circle), Cu$X$s in the BSQ structure have positive $E_{j}$ and those in the B2 and L1$_0$ structures are energetically more stable. The B2 and/or L1$_0$ structures have negative $E_{j}$ for the cases of $X=$ Li, Be, Mg, Ca, Sc, Y, Lu, Ti, Zr, Pd, Pt, Au, Zn, Al, and Ga. The group-dependence of the energetic stability is similar to that in the B$_h$ and L1$_1$ structures and that predicted for the Pb- and Sn-based compounds \cite{ono2021_CPL}.  

In the present study, Cu$X$ in the B2 structure for $X=$ group 1 (K and Rb), group 2 (Be, Mg, Ca, Sr, and Ba), group 3 (Sc, Y, and Lu), and Pd have been identified to be dynamically stable. In contrast, CuBe \cite{CuBe}, CuSc \cite{CuSc}, CuY \cite{CuY}, CuZr \cite{CuZr}, CuPd \cite{CuPd}, and CuZn \cite{CuZn} in the B2 structure have been synthesized experimentally at ambient condition. For the case of the L1$_0$ structure, 20 Cu$X$ in the L1$_0$ structure have been predicted to be dynamically stable, whereas only CuAu has been synthesized experimentally \cite{CuAu}. It has been shown that the lowest energy phonon at the R point, $(0,\pi/a,\pi/c)$, of the L1$_0$ CuAu is stabilized by the long-range interatomic interactions from the first to more than the fifth nearest neighbors \cite{ono2021_wdm}. If the same scenario holds, the stabilization of the phonons at the N point, $(0,\pi/a,\pi/a)$, and the R point will be a key to synthesize the ordered compounds in the B2 and L1$_0$ structures, respectively. 

\subsection{Tetragonal and trigonal paths}
\label{sec:bain_trig}
\begin{figure*}[tt]
\center
\includegraphics[scale=0.45]{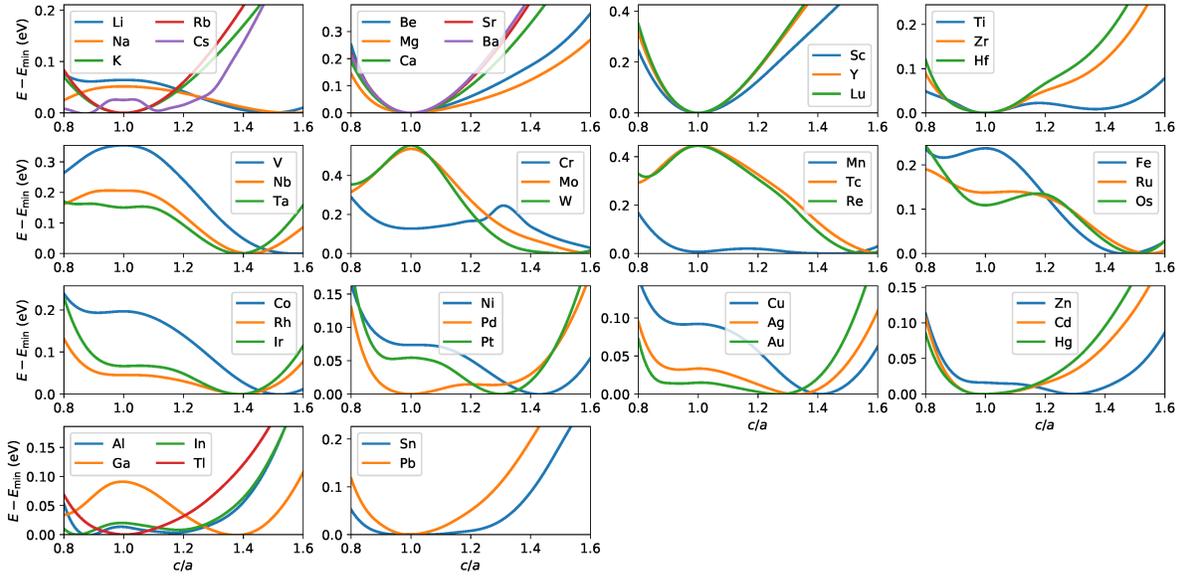}
\caption{The total energy per a cell as a function of $c/a$ (the tetragonal path) for Cu$X$. } \label{fig_bain} 
\end{figure*}

To understand the metastability relationship between the B2 and the L1$_0$ structures proposed in Sec.~\ref{sec:BSQ_B2_L1_0}, we show the total energy curves along the tetragonal Bain path in Fig.~\ref{fig_bain}, where the volume of Cu$X$ is fixed to that of the B2 structure. If the total energy takes the minimum value at $c/a=1$ (B2), no energy minimum is observed at $c/a>1$ (L1$_0$), as for $X=$ K, Rb, group 2, and group 3. In contrast, if the total energy takes the minimum value at $c/a>1$, then that takes the maximum or higher values at $c/a=1$. It should be noted that the anomalous stability of CuPd in the B2 structure (see the group 10 in Fig.~\ref{fig1}) can be understood as the shift of the energy minimum to $c/a=1$.

\begin{figure*}[tt]
\center
\includegraphics[scale=0.45]{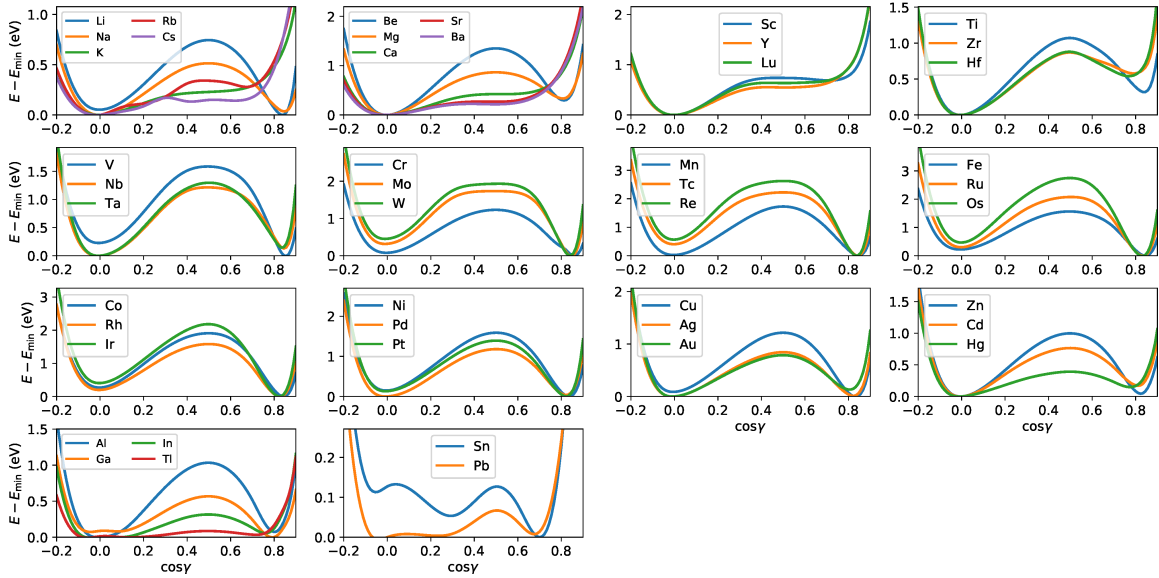}
\caption{The total energy per a cell as a function of ${\rm cos}\gamma$ (the trigonal path) for Cu$X$. } \label{fig_trig} 
\end{figure*}

The B2 structure can be transformed into the L1$_1$ structure through the trigonal path, elongating or shortening the cubic structure along the (111) direction with the volume fixed to that of the B2 structure. Figure \ref{fig_trig} shows the total energy as a function of $\cos\gamma$, where $\gamma$ is the angle between the primitive lattice vectors. For most of $X$, the double well-like energy curves are observed, indicating that the B2 ($\cos\gamma=0$) and L1$_1$ ($\cos\gamma\simeq 0.8$) structures are energetically stable. The energy maximum at $\cos\gamma=0.5$ corresponds to the B1 structure. For $X=$ Ca, Sr, Ba, Sc, Y, and Lu, the energy curve around $\cos\gamma=0.5$ is almost flat. In these compounds, the L1$_1$ structure was transformed into the B1 structure after the geometry optimization (see the SM \cite{SM}). 

The stability properties of Cu$X$ along the tetragonal and trigonal paths in Figs.~\ref{fig_bain} and \ref{fig_trig} are almost consistent with those in Figs.~\ref{fig1} and \ref{fig2}. However, the energetic stability along the tetragonal and trigonal paths may not guarantee the dynamical stability of the structure \cite{grimvall}. For $X=$ Ti, Zr, and Hf (group 4), although the trigonal path calculations predict that the B2 and L1$_1$ structures are stable, Cu$X$ in those structures are unstable (see Figs.~\ref{fig1} and \ref{fig2}). This implies that other transformation paths may exist, leading to a negative curvature in the potential energy surface around the B2 and L1$_1$ structures. It will be desirable to perform more systematic calculations that investigate the interrelationship between metastable structures, as demonstrated in elemental metals \cite{togo}. 


\subsection{CuAu in the B$_h$ and L1$_1$ structures}
\label{sec:cuau}
The experimental synthesis of the BHC CuAu has been reported recently \cite{zagler}. Based on the phonon calculations, we confirmed that such a structure is dynamically stable \cite{ono2021} and predicted that the B$_h$ and L1$_1$ structures are also dynamically stable, as they are constructed from the 2D structures. We discuss the temperature effect on the stability in the 3D structures below.  

For the B$_h$ structure, the lattice parameters are $a=2.858$ \AA \ and $c/a=1.526$. For the L1$_1$ structure, the lattice parameter is $a=4.726$ \AA \ and the primitive vectors are $a(u,v,v)$, $a(v,u,v)$, and $a(v,v,u)$ with $u=0.239$ and $v=0.669$. The total energy of the B$_h$ and L1$_1$ structures are energetically higher than that of the L1$_0$ structure, the ground state structure of CuAu, by 0.176 eV and 0.138 eV per a cell. These values are much higher than the thermal energy at the ambient condition. As shown in Fig.~\ref{fig_trig}, the potential barrier height ($\sim 1$ eV) between the L1$_1$ and the B2 ($\simeq$ L1$_0$) structures is also high enough not to yield the phase transition. Therefore, once synthesized, these metastable structures will be stable. 

\begin{figure}[tt]
\center
\includegraphics[scale=0.35]{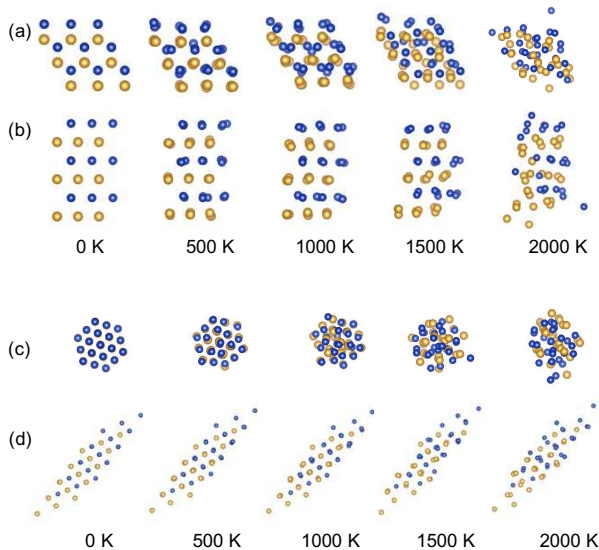}
\caption{The atomic distribution of the CuAu compounds after the MD simulation of 1 ps for several $T$s: (a) Top and (b) side views of the B$_h$ structure and (c) top and (b) side views of the L1$_1$ structure. } \label{fig_Bh_MD} 
\end{figure}

Figure \ref{fig_Bh_MD}(a) and \ref{fig_Bh_MD}(b) shows the temperature ($T$)-dependence of the atomic distribution in the B$_h$ supercell at the end of the molecular dynamics (MD) simulation (1 ps). The B$_h$ structure is stable up to $T=1000$ K because the structure of each layer (i.e., the 2D hexagonal layer of Au and Cu) is still preserved. When $T=1500$ K, the hexagonal symmetry in each layer is broken. When $T$ is increased to 2000 K, the atomic displacements are significant, so that the layered structure is not observed. Similar tendency is observed in the L1$_1$ structure, as shown in Fig \ref{fig_Bh_MD}(c) and \ref{fig_Bh_MD}(d). 

\section{Conclusion}
In conclusion, by focusing on Cu$X$, we have investigated the metastability relationship between the 2D and 3D structures. We have demonstrated that (i) if the BHC structure is dynamically stable, the B$_h$ and L1$_1$ structures are also stable, as summarized in Table~\ref{table1}. As an example, we have shown that CuAu in these structures are stable up to 1500 K. We have also established that (ii) if Cu$X$ in the B2 (L1$_0$) structure is dynamically stable, that in the L1$_0$ (B2) structure is unstable, which is analogous to the metastability relationship between the fcc and the bcc structures in elemental metals. We hope that the present study paves the way to computational materials design based on the interrelationship between the 2D and 3D structures. Making the metastability database for other compounds $XY$ and exploring other 2D structures as building blocks for the 3D structures are left for future work. 


\begin{acknowledgments}
This study was supported by the a Grant-in-Aid for Scientific Research (C) (Grant No. 21K04628) from JSPS. The computation was carried out using the facilities of the Supercomputer Center, the Institute for Solid State Physics, the University of Tokyo. 
\end{acknowledgments}


\appendix

\section{Computational details}
The total energy and the optimized structures of Cu$X$ were calculated within the density-functional theory implemented in \texttt{Quantum ESPRESSO} (\texttt{QE}) \cite{qe}. We used the exchange-correlation functional of Perdew, Burke, and Ernzerhof (PBE) \cite{pbe} and the ultrasoft pseudopotentials in \texttt{pslibrary.1.0.0} \cite{dalcorso}. We used the cutoff energies of 80 and 800 Ry for the wavefunction and the charge density, respectively, used 20$\times$20$\times$1 $k$ grid and 20$\times$20$\times$20 $k$ grid for 2D and 3D structures, respectively \cite{MK}, and set the smearing parameter of $\sigma=0.02$ Ry \cite{smearing}. For the BHC and BSQ structures, the size of the unit cell along the $c$ axis was set to be 14 \AA \ to avoid the spurious interaction between different unit cells. For the geometry optimization of the BHC, B$_h$, BSQ, B2, and L1$_0$ structures, the initial guess of the lattice parameters was set to the same as the optimized value obtained by our previous calculations \cite{ono2021}. The optimization calculations of the L1$_1$ structure was started by assuming $\cos\gamma=0.82$ ($\gamma=35^\circ$ between the primitive lattice vectors) and the unit cell volume equal to that of the B$_h$ structure.  

We performed the spin-polarized calculations, where the total energy and forces are converged within $10^{-5}$ Ry and $10^{-4}$ a.u., respectively, which are smaller than the default values by a factor of ten. In our previous study, we performed the spin-unpolarized calculations by using the default values of the convergence criteria \cite{ono2021}. In the spin-polarized calculations, the finite magnetic moment was observed for Cu$X$ with $X=$ Cr, Mn, Fe, Co, and Ni when a ferromagnetic state was assumed as an initial guess. As listed in Table \ref{table3}, the size of the magnetic moment in the BHC and BSQ structures seems to be larger than that in their 3D counterparts. We also found that a few compounds show a finite magnetic moment: For the BHC structure, CuTi (0.87); for the B2 structure, CuHg (0.47), CuIr (0.47), CuRe (2.28), CuRh (1.02), and CuZr (0.01); and for the L1$_0$ structure, CuHf (0.01), CuRh (0.72), and CuZr (0.05), where the figure in parenthesis is in units of $\mu_{\rm B}$ per a cell. It has been shown that the 2D elemental metals of Ba, group 2 (Sc and Y), group 3 (Ti, Zr, and Hf), V, group 8 (Ru and Os), group 9 (Rh and Ir), and group 10 (Pd and Pt) can have a finite magnetic moment \cite{ren}. This has been attributed to a decrease in the coordination number as well as a change in the electron DOS shape. Therefore, it will be helpful to study the effect of the electronic band structure on the dynamical stability of Cu$X$, in addition to the geometric concept investigated in the present work. It will also be interesting to study the competition of the energetic stability between the ferromagnetic and antiferromagnetic phases in a systematic manner. However, these are beyond the scope of the present work.  

\begin{table}
\begin{center}
\caption{The total magnetic moment of Cu$X$ ($X=$ Cr, Mn, Fe, Co, and Ni) in the structure $j$. The figures are in units of $\mu_{\rm B}$ per a cell ($\mu_{\rm B}$ the Bohr magneton). }
{
\begin{tabular}{lcccccc} \hline\hline
$X$ & \hspace{2mm} BHC \hspace{2mm} & \hspace{2mm} B$_h$ \hspace{2mm} & \hspace{2mm} L1$_1$ \hspace{2mm} & \hspace{2mm} BSQ \hspace{2mm} & \hspace{2mm} B2 \hspace{2mm} & \hspace{2mm} L1$_0$ \hspace{2mm} \\
\hline
Cr & 0.00 & 0.00 & 0.00 & 0.00 & 3.47 & 0.00 \\
Mn & 3.83 & 2.98 & 0.00 & 0.00 & 0.00 & 0.00 \\
Fe & 2.64 & 2.55 & 2.54 & 2.80 & 2.66 & 2.71 \\
Co & 1.70 & 1.67 & 1.65 & 1.84 & 1.53 & 1.74 \\
Ni & 0.47 & 0.30 & 0.00 & 0.55 & 0.41 & 0.43 \\
\hline\hline
\end{tabular}
}
\label{table3}
\end{center}
\end{table}

\begin{table}
\begin{center}
\caption{The lattice parameters of Cu$X$ in the B2 structure, CuAu in the L1$_0$ structure, and CuPt in the L1$_1$ structure. For the L1$_1$ structure, the parameters of the conventional unit cell (including six atoms) are listed. }
{
\begin{tabular}{lccr} \hline\hline
$X$ & \hspace{2mm} this work \hspace{2mm}  & \hspace{2mm} MP \cite{MP} \hspace{2mm} & \hspace{5mm} Experiment \\
\hline
Be & 2.69 & 2.69 & 2.70 \cite{CuBe} \\
Pd & 3.01 & 3.02 & 2.98 \cite{CuPd} \\
Sc & 3.26 & 3.26 & 3.26 \cite{CuSc} \\
Y  & 3.48 & 3.48 & 3.48 \cite{CuY} \\
Zn & 2.96 & 2.96 & 2.95 \cite{CuZn} \\
Zr & 3.27 & 3.27 & 3.26 \cite{CuZr} \\
Au ($a$) & 2.87 & 2.86 & 2.85 \cite{CuAu} \\
Au ($c$) & 3.64  & 3.66 & 3.67 \cite{CuAu} \\
Pt ($a$) & 2.87 & 2.75 & 3.13 \cite{CuPt} \\
Pt ($c$) & 12.91  & 12.86 & 14.98 \cite{CuPt} \\
\hline\hline
\end{tabular}
}
\label{table2}
\end{center}
\end{table}

The formation energy of Cu$X$ in the structure $j$ per a unit cell including two atoms was calculated by 
\begin{eqnarray}
 E_{j}({\rm Cu}X) = \varepsilon_{j}({\rm Cu}X) 
 - \frac{1}{2} \left[ \varepsilon_{\rm min}({\rm Cu}) + \varepsilon_{\rm min}(X) \right],
 \label{eq:Eform}
\end{eqnarray}
where $\varepsilon_{j}({\rm Cu}X)$ is the total energy of Cu$X$ in the structure $j$. $\varepsilon_{\rm min}(X)$ is the minimum total energy of $X$, where this is determined by calculating the energy of the bcc, fcc, and hcp structures. Note that the functional dependence of the $E_j$ of ordered compounds has been studied in detail \cite{nonlocalpbe,isaacs,ruzsinszky2019,ruzsinszky2020}. The use of the PBE functional can predict the value of $E_j$ accurately for strongly bonded systems, while it may underestimate $E_j$ for weakly bonded systems such as CuAu. The calculated values of $E_j$ as well as those in the Materials Project (MP) database \cite{MP} were provided in the SM \cite{SM}. 

To show the numerical accuracy within the PBE calculations, we compared the lattice parameters with those in the MP database \cite{MP}, and the experimental values for the B2, L1$_0$, and L1$_1$ structures, as listed in Table~\ref{table2}. Except for CuPt, the agreement between the present calculations, the MP database \cite{MP}, and the experiment is good (the deviation is less $\pm$ 0.03 \AA). The lattice parameters of Cu$X$ in the structure $j$ were also provided in the SM \cite{SM}. 

The dynamical stability and the instability of Cu$X$ in the structure $j$ were determined by the phonon DOS calculations within the density-functional perturbation theory \cite{dfpt} implemented in the \texttt{QE} code \cite{qe}. We used more than a 6$\times$6$\times$1 $q$ grid for the 2D structures (seven $q$ points for the BHC and ten $q$ points for the BSQ structures), 3$\times$3$\times$3 $q$ grid (six $q$ points) for the B$_h$ and L1$_0$ structures, 3$\times$3$\times$3 $q$ grid (four $q$ points) for the B2 structure, and 2$\times$2$\times$2 $q$ grid (four $q$ points) for the L1$_1$ structure.


For the case of CuAu in the B$_h$ and L1$_1$ structures, the MD simulation was performed by using the \texttt{QE} \cite{qe}. The 3$\times$3$\times$3 supercell (54 atoms) was used by considering only the $\Gamma$ point in the Brillouin zone. The ionic temperature $T$ is controlled by the velocity scaling method. The time step of the MD simulation was set to 20 a.u. (0.96 fs) and the number of the steps was set to 1050 (1.01 ps) with $T=500$, 1000, 1500, and 2000 K.  


\end{document}